\documentclass[twocolumn,usenatbib,useAMS]{mnras}
\PassOptionsToPackage{dvipsnames,svgnames,x11names}{xcolor}
\usepackage{xparse,soul}

\makeatletter
\NewDocumentCommand{\sotwo}{O{red}O{black}+m}
    {
        \begingroup
        \color{#1}%
        \setul{-.5ex}{.4pt}%
        \aef\SOUL@uleverysyllable{%
            \rlap{%
                \color{#2}\the\SOUL@syllable
                \SOUL@setkern\SOUL@charkern}%
            \SOUL@ulunderline{%
                \phantom{\the\SOUL@syllable}}%
        }%
        \ul{#3}%
        \endgroup
    }
\makeatother

\usepackage{graphicx}	
\usepackage{amsmath}	
\usepackage{amssymb}	
\usepackage{multicol} 
\usepackage{multirow}
\usepackage{bm}	
\usepackage{pdflscape}
\usepackage{caption}
\usepackage{subfigure}
\usepackage{siunitx}
\usepackage{color}
\usepackage{lastpage}
\usepackage{xspace}

\newcommand{\kms}{\,km\,s$^{-1}$} 

\newcommand{\simba}{\mbox{{\sc Simba}}\xspace}
\def\bsimba{\textbf{\textsc{Simba}}}
\newcommand{\hMpc}{\,h^{-1} \, {\rm cMpc}}
\usepackage[T1]{fontenc}
\usepackage{ae,aecompl}
\newcommand{\specialcell}[2][c]{%
\begin{tabular}[#1]{@{}c@{}}#2\end{tabular}}
\usepackage{newtxtext,newtxmath}

\title[tSZ in \simba and TNG]{Understanding the relation between thermal Sunyaev-Zeldovich decrement and halo mass using the \bsimba\ and TNG simulations}

\author[Yang et al.]{Tianyi Yang$^{1}$, Yan-Chuan Cai$^{1}$, Weiguang Cui$^{1,2}$, Romeel Dav\'e$^{1,3,4}$, John A. Peacock$^1$,
\newauthor and Daniele Sorini$^{1}$
\\
\\$^1$ Institute for Astronomy, University of Edinburgh, Royal Observatory, Blackford Hill, Edinburgh EH9 3HJ, UK
\\$^2$ Departamento de Física Teórica and CIAFF, Modulo 8 Universidad Autónoma de Madrid, 28049 Madrid, Spain.
\\$^{3}$ University of the Western Cape, Bellville, Cape Town 7535, South Africa
\\$^{4}$ South African Astronomical Observatories, Observatory, Cape Town 7925, South Africa
}

\pubyear{2022}

\begin{document}
\label{firstpage}
\pagerange{\pageref{firstpage}--\pageref{lastpage}}
\maketitle

\begin{abstract}
The relation between the integrated thermal Sunyaev-Zeldovich (tSZ) $y$-decrement versus halo mass ($Y$--$M$) can potentially constrain galaxy formation models, if theoretical and observational systematics can be properly assessed. We investigate the $Y$--$M$ relation in the \simba and IllustrisTNG-100 cosmological hydrodynamic simulations, quantifying the effects of feedback, line-of-sight projection, and beam convolution. We find that \simba's AGN jet feedback generates strong deviations from self-similar expectations for the $Y$--$M$ relation, especially at $M_{\rm 500}\la 10^{13}M_\odot$. In \simba, this is driven by suppressed in-halo $y$ contributions owing to lowered halo baryon fractions. IllustrisTNG results more closely resemble \simba without jets. Projections of line-of-sight structures weaken these model differences slightly, but they remain significant -- particularly at group and lower halo masses.  In contrast, beam smearing at {\it Planck\/} resolution makes the models indistinguishable, and both models appear to agree well with {\it Planck\/} data down to the lowest masses probed.  
We show that the arcminute resolution expected from forthcoming facilities would retain the differences between model predictions, and thereby provide strong constraints on AGN feedback. 

\end{abstract}


\section{Introduction}\label{sec::intro}

Galactic haloes are not closed boxes. They grow via gravitational accretion from the intergalactic medium (IGM), which is expected to occur with a baryon-to-dark matter ratio close to the cosmic mean. However, haloes can also lose baryons through energetic feedback processes \citep[e.g.][]{Tollet_2019, Appleby_2021, Lim_tng_eagle_mag_sim_paper, Mitchell_2021, Sorini_2021, Gas_distribution_from_clusters_to_filaments_in_IllustrisTNG}, leaving a deficit of baryons in the halo relative to the cosmic mean. In haloes smaller than galaxy clusters, this {\it missing halo baryon problem\/} is established, in that the amount of baryons that can be robustly observed does not sum up to the expectations from the mean cosmic baryon fraction ~\citep[e.g.][]{missing_baryon_paper_ref1_Mcgaugh_2000,missing_baryon_paper_ref1_Mcgaugh_2010,Tumlinson_CGM_review}. It remains unclear whether this indicates that a fraction of the baryons remain undetected, or that the missing baryons have been ejected from the halo altogether.

Around massive galaxies, the baryons within the circum-galactic medium (CGM, broadly defined here as the gas within the halo) and the surrounding IGM are thought to be predominantly hot ($T\ga 10^6$K), which means they can in principle be detected via X-rays. However, X-ray emission is generated by collisional processes that scale as the density squared, meaning that such observations tend to be better probes of the inner high-density regions of haloes rather than the outskirts or the IGM. Nonetheless, X-ray measurements suggest deficiencies in the halo baryon fractions emerging at group scales~\citep[e.g.][]{McCarthy_bayonic_fraction_BAHAMAS}.  This problem further extends to Milky Way sized halo masses ($M_{\rm halo}\sim 10^{12}M_\odot$), which also may show a substantial missing halo baryon problem~\citep[e.g.][]{Peeples_CGM_bayonic_fraction}, albeit with large uncertainties from absorption line probes.  The increasingly multi-phase CGM and the diffuse nature of the gas thus makes a halo baryon census highly challenging at galaxy group scales.

Models and simulations of galaxy formation are now routinely successful at reproducing the populations of galaxies, such as stellar mass functions or the properties of gas and metals within galaxies~\citep[e.g.][and references therein]{somerville_dave_2015_galaxy_formation_model}. However, the gas properties around galaxies have not typically been used to constrain models. To have a full picture of galaxy formation, it is important to test models critically against the gas properties around haloes, which may provide strong complementary constraints on the physical processes of galaxy evolution.

An emerging probe of CGM baryons is provided by the thermal Sunyaev-Zel’dovich (tSZ) effect~\citep{SZ_original_paper}, an energy shift in Cosmic Microwave Background (CMB) photons caused by the inverse Compton scattering of the  photons by hot electrons within haloes or the surrounding IGM, yielding a frequency shift to the original CMB radiation spectrum. The magnitude of this shift is characterised by the Compton parameter $y$, a dimensionless quantity that is related to the integral of gas pressure along the line of sight to the surface of last scattering:
\begin{equation}\label{eqn:y_params_eq}
y=\frac{\sigma_{\rm T}}{m_{e}c^2}\int P_{e}\;d\ell \propto \int n_{e} T_{e}\;d\ell,   
\end{equation}
where $\sigma_{\rm T}$ is the cross-section of electron Thomson scattering and $m_{e}c^2$ is the electron rest energy, $P_{e}$ is the electron pressure calculated as $n_{e}k_{\rm B}T_{e}$ with $k_{\rm B}$ being the Boltzmann constant, $n_e$ and $T_{e}$ being the electron number density and temperature respectively. This quantity directly probes the thermal energy of the Universe, and compared to other probes (such as X-rays) it is less sensitive to the gas density,
and is independent of redshift. 

The tSZ effect corresponds to a change in CMB surface brightness in a given direction that is proportional to (hereafter $\propto$) $y$. A convenient total observable is then the flux density given by integrating this surface brightness over angle on the sky:
\begin{equation}
    Y \equiv \int y({\boldsymbol\theta})\; d^2\theta.
\end{equation}
This observable depends on the distance to the system: $y$ itself is independent of distance, but the angular size of a halo will be smaller at high redshift. Thus we might expect a distant-independent intrinsic property of a cluster to be $Y\, D_{\rm A}^2$ ~$\propto$~$\int n_e T\, dV$, where $D_{\rm A}$ is the proper angular-diameter distance.

In practice, however, it is common to choose a slightly different expected scaling of $Y$ with redshift. This derives from the idea of {\it self-similar\/} cluster evolution \citep[e.g.][]{1998ApJ...495...80B, silva_paper_with_scaling_rel, scaling_relations_for_clusters_review}. Here one assumes that clusters have universal density profiles, with a proper virial radius $R$, within which the density is some multiple of the mean, $\Delta\bar\rho$, so that $M$~$\propto$~$\Delta\bar\rho(z) R^3$. The virial temperature should scale as $T$~$\propto$~$GM/R$, so that
\begin{equation}
Y \, D_{\rm A}^2 \propto f_{\rm gas} \Delta^2\bar\rho^2 R^5
\propto f_{\rm gas}\, M^{5/3}\, (\Delta\bar\rho)^{1/3},
\end{equation}
where we also allow for an arbitrary mass fraction of hot gas. If the overdensity $\Delta$ were a constant, this would predict that a given system placed at high redshift would display a tSZ signal that is stronger by a factor of $(1+z)$. An alternative convention, which we follow here, is that the virial radius is set at some multiple of the {\it critical\/} density, so that $\Delta\bar\rho$~$\propto$~$ \rho_c(z)$~$\propto$~$H(z)^2$. Absorbing this evolution leads to the following definition of a rescaled tSZ flux:
\begin{equation}
    \bar Y \equiv E(z)^{-2/3} D_{\rm A}^2 Y,
\end{equation}
where $E(z) = H(z)/H_{0} = \sqrt{\Omega_{m}(1+z)^{3}+\Omega_{\Lambda}}$. Due to the evolving critical density of the Universe, a tSZ signal ($YD_{A}^2$) at high redshift is stronger by a factor of $\smash{E(z)^{2/3}}$. Therefore by design, $\bar{Y}$ should be redshift independent. 

A key test of this hierarchical evolution is to see if the predicted relation, $\bar Y$~$\propto$~$M^{5/3}$, is obeyed. In principle, we should work with $M_{\rm halo}$, the total halo mass, but for consistency with other studies, we use $M_{500}$ as a proxy for the halo mass: $M_{500}$ is defined as the mass contained by a sphere of radius $r_{500}$, within which the mean density is 500 times the critical value. Thus our empirical tSZ probe of the total gas pressure in and around haloes will be the $\bar Y-M_{500}$ relation. Deviations from the self-similar model prediction would indicate departures from the simple model assumptions, which can be sensitive to the presence of non-gravitational energy input e.g., from AGN feedback \citep[e.g. in][]{Le_brun_2017, Barnes_2017,Truong_2018}\footnote{As discussed by these papers, the departures from self-similarity can be driven by other factors such as non-thermal pressure and spectroscopic temperature bias from X-ray observations.}. As tSZ observations improve, this opens up the opportunity of using such data to constrain processes of galaxy evolution. In fact, the possibility of using SZ observations to study baryonic physics in group-sized haloes has been testified by several recent studies \citep{ACT_BOSS_paper1,ACT_BOSS_paper2,ACT_BOSS_paper3, Moser_2022}.

The release of the $\textit{Planck}$ all-sky $y$-map \citep{planck_NILC,planck_MILCA,planck_tsz_paper} allows observers to probe the tSZ signal around haloes down to near Milky Way-like halo masses via stacking. \citet{Planck_Y_M_paper} measured $Y$ for a set of locally brightest galaxies as a function of their halo masses $M_{500}$, where the halo masses of these galaxies were estimated using a specific central galaxy stellar mass--halo mass relation. Their results suggested that the self-similar relation between $Y$ and $M_{500}$ was satisfied down to $\sim 10^{12.5} M_{\odot}$. At face value, this indicates that non-gravitational effects are mild even down to Local Group-sized haloes, and that such poor groups contain close to their cosmic share of baryons.

Studies from simulations and observations for the $Y$--$M_{\rm 500}$ relation seemed to yield divergent conclusions. At cluster mass scales, \citet{X_ray_surface_brightness_form_and_some_good_exp} and \citet{Nagai_SZ_sim_paper} showed that this relation predicted from simulations was well-described by the self-similar model, but their simulations did not include AGN feedback. \citet{Arnaud_paper} recovered a self-similar model between $Y$ and $M_{500}$ using X-ray observations for $10^{14} M_{\odot}$ to $10^{15} M_{\odot}$ haloes, where feedback effects are expected to be minor compared to gravitational heating. Using the dataset analyzed by \citet{Planck_Y_M_paper}, \citet{Planck_Y_M_contin} recovered the self-similar model by studying the tSZ signal around locally brightest galaxies with $ \textrm{log}_{10}(M_{\ast}/M_{\odot})>11.3$. Compared to the {\it Planck\/} team, they corrected for contamination from dust emission and from foreground and/or background objects in a more explicit way. Their findings were consistent with the {\it Planck\/} observations. Using hydrodynamic simulations of $\sim 300$ galaxy clusters with and without AGN feedback, \cite{pymsz_ref_paper, Cui2022} found a $Y$--$M_{\rm 500}$ relation that was also very close to self-similar with different baryon models. Note that their result is mostly from the massive galaxy clusters. However, \citet{Barnes_2017} found a steeper $Y$--$M_{\rm 500}$ relation using galaxy clusters with $M_{500} \geq 10^{14} M_\odot$ selected from combined MACSIS and BAHAMAS simulations at $z = 0$, and also a mild evolution with redshift. A similar slope for clusters in $10^{14} \leq M_{500} \leq 10^{15.5}$ is also been reported in \citet{Le_brun_2017} with their `AGN 8.5' model. As pointed out by \cite{Fable_Y_x_M500}, this slope is halo mass dependent -- for their most massive sample, the slope of the $Y_{500}$--$M_{500}$ relation is consistent with the
self-similar value. Recently, by measuring the stacked tSZ signal around DESI galaxy clusters/groups using the Planck tSZ $y$-map, \citet{new_Y_M_paper} found a slightly steeper $Y$--$M$ slope than the self-similar model, which also had a slight tendency to increase with redshift.

However, predictions of this relation from hydrodynamic simulations for smaller group-scale haloes deviate from self-similarity. \citet{silva_paper_with_scaling_rel} found that when cooling and heating processes were included in the simulations, a broken power law was required for an accurate description of the measured $Y$--$M_{\rm 500}$ relation at both the low-mass and high-mass ends. \citet{Lim_tng_eagle_mag_sim_paper} reported that when using IllustrisTNG \citep{tng_AGN_feedback, tng_sim_ref1, tng_sim_ref2}, EAGLE \citep{eagle_sima_ref1,eagle_sim_ref2} and Magneticum simulation data \citep{mag_sim_paper}, AGN feedback resulted in $Y$ values for group-sized haloes that were significantly lower than the self-similar prediction. 

\citet{Le_Brun_paper} used synthetic tSZ maps from simulations to point out that the observed $Y$--$M_{\rm 500}$ relationship is sensitive to the assumed pressure distribution of the gas, and that a model with strong AGN feedback seemed to give a better match with the {\it Planck\/} observations. AGN feedback is now regarded as a necessary component of galaxy formation models in order to produce a realistic galaxy population \citep[e.g.][and references therein]{somerville_dave_2015_galaxy_formation_model}, but the amount of energy needed and the way the energy is fed back into the gas remains unclear.  Ongoing and future high-resolution high-sensitivity tSZ observations, such as NIKA2 \citep{NIKA2}, the Simons Observatory \citep{SO_ref}, and CMB-S4 \citep{CMB_S4_ref}, will provide a tightly constrained $Y$--$M_{500}$ relation at these mass scales, which if interpreted properly could be used to constrain the AGN feedback models.

Facing this ambiguous situation for the $Y$--$M_{\rm 500}$ relation, here we approach the problem by carefully considering a number of challenges in the $Y$--$M_{\rm 500}$ analysis: (i) the observed $Y$ receives contributions of all gas along the line of sight to the CMB, and the gas pressure from correlated and uncorrelated large-scale structure will confuse the interpretation of $Y$ around haloes; and (ii) the smoothing of the observed $y$-map due to the finite resolution of observations may wash out important features of the signal at small scales, obscuring the underlying differences between models. A focus of our study is to use simulations to understand some of these contaminants in observations.

In this paper, we examine the $Y$--$M_{500}$ relation in the $\simba$ and IllustrisTNG simulations. We create mock {\it Planck\/} $y$ maps from these simulations, and measure the resulting $Y$ values in a manner similar to the procedure used by the {\it Planck\/} team. We first compare the resulting $\textit{true}$ $Y$ versus the contribution from gas within $r_{500}$ in 3D and in 2D within the full projected area of the halo, thereby quantifying how much of the tSZ signal comes from the foreground and background two-halo terms. We then use the $\simba$ simulation's variant runs that turn off specific feedback processes in order to pinpoint which have the greatest impact on the tSZ signal. We examine gas densities and temperatures in $\simba$ giving rise to the line-of-sight pressure, in order to see more clearly how feedback processes change $y$. Finally, we perform similar analyses as in observations with our simulated maps, explore whether future CMB experiments could distinguish between feedback models. Overall, we show that all simulations are consistent with the {\it Planck\/} data in terms of $Y$--$M_{500}$ relations, owing to the convolution with the 10-arcmin {\it Planck\/} beam. However, they show substantial differences when we create higher resolution tSZ maps, and would be distinguishable with future tSZ observatories.

The layout of this paper is as follows. In Section \ref{sec::simulation_data}, we briefly introduce some key properties of the $\simba$ and TNG projects, and  discuss the main differences of the AGN feedback models employed by these two simulations. In Section \ref{sec::reconstrunction_of_y_map}, we demonstrate the reconstructed tSZ $y$ maps and discuss the gas distribution around haloes obtained from different simulation runs. In Section \ref{sec::Y_M_relation}, we present the resulting $Y$--$M$ relations measured from simulations and compare them with observations. We conclude with a discussion of our findings in Section \ref{sec::conclusion}.

\section{Cosmological Simulations}\label{sec::simulation_data}

The goal of our study is to confront predictions from cosmological hydrodynamic simulations with tSZ observations. Since state-of-the-art hydro-simulations are typically tuned to reproduce the observational properties of galaxies, our focus here is to study hot gas properties predicted from these simulations, and see if the differences between these predictions are large enough to allow these models to be distinguished.

Our analysis is based on two suites of hydrodynamic simulations. We primarily focus on $\simba$ \citep{simba_ref}, but for comparison we also consider IllustrisTNG100-2 \citep[hereafter TNG100-2;][]{tng_sim_ref2} which has a similar resolution to $\simba$. Some basic information is summarised in Table \ref{tab:sim_info}. For galaxy evolution, these simulations adopt different input gas physics, black hole growth and feedback models. For $\simba$ we will use both the main full-physics $(100 \hMpc)^{3}$ run, as well as `feedback variant' runs performed in a $(50\hMpc)^{3}$ box with identical input physics, with the same resolution as the main run but with one-eighth the particles. The feedback variants allow us to isolate the impact of specific feedback processes on the gas distribution around haloes. 

\begin{table}
    \centering
     \caption{Some relevant information regarding the simulations used in this work. From left-to-right: simulation name, comoving box size ($\hMpc$), initial mean gas cell resolution ($M_{\odot}$), Dark matter particle mass resolution ($M_{\odot}$).}
    \begin{tabular}{c|ccc}
    \hline
         & box size & cell resolution & particle mass\\
         \hline
         $\simba$ &  \specialcell{50 (small) \\100 (large)} & $1.82\times10^{7}$ &  $9.58\times10^{7}$\\
         \hline
         IllustrisTNG100-2 & 75 &  $1.11\times10^{7}$&  $5.97\times10^{7}$\\
         \hline
    \end{tabular}
    \label{tab:sim_info}
\end{table}

\subsection{The \bsimba\ simulations}\label{ssec::simba_intro}

$\simba$\footnote{Snapshots and catalogues available at \url{http://simba.roe.ac.uk/}} \citep{simba_ref} is a suite of hydrodynamic simulation using the $\textsc{Gizmo}$ code \citep{Gizmo_ref}. Dark matter and gas particles are evolved within a periodic cubical volume with a $\textit{Planck}$ 2015 concordance cosmology \citep{planck_2015_cosmo_paper} of $\Omega_{m} = 0.3, \Omega_{\Lambda} = 0.7, \Omega_{b} = 0.048, H_{0} = 68 \rm ~km~s^{-1}~Mpc^{-1}, \sigma_{8} = 0.82$ and $n_{s} = 0.97$. The primary run has a box length of 100 \textrm{comoving $h^{-1}$\,~Mpc} (hereafter $h^{-1}$\,cMpc) (denoted \simba-100). To test sensitivity to input physics, there are several 50 $\hMpc$ boxes (denoted \simba-50).  For these, $\simba$ includes different runs with different feedback models turned on/off, as described below. The initial conditions for all \simba-50 runs are identical.

$\simba$ includes a wide range of input physics designed to reproduce galaxies as observed across cosmic time.  These include radiative cooling and heating, star formation, stellar evolution, feedback associated with young stars and supernovae, and the formation and evolution of dust.  Specifics of these models are available in \citet{simba_ref}.

Owing to its importance for this work, we describe the black hole growth and active galactic nuclei (AGN) feedback in $\simba$ in more detail. Black hole growth is simulated via two distinct modes: cold gas ($T<10^5$) is accreted onto the black hole following the torque-limited accretion model of \citet{torque_limited_accretion}, while hot gas is accreted according to Bondi capture \citep{bondi_accretion_ref}. To model AGN feedback, energy is released into the surrounding region via kinetic outflows \citep{feedback_mechanism_ref,torque_limited_accretion}. The velocity of the outflows depends on the ratio of the accretion rate to the Eddington rate mimicking \citet{feedback_mechanism_ref}, with high ratios yielding `radiative mode' feedback at $v\sim 500-1500$\kms and low ratios yielding `jet mode' feedback with ejection velocities approaching $10^4$\kms.  Outflows are stably bipolar, ejected parallel/anti-parallel to the angular momentum vector of the 256 closest gas particles (typically $\sim 1$\,kpc).  When jet mode is active, an additional X-ray feedback mechanism is included which mimics the deposition of high-energy photons into surrounding gas.  

These feedback mechanisms work together to yield good agreement versus many observables, including the stellar mass function evolution and the fraction of quenched galaxies \citep{simba_ref}; the black hole--galaxy scaling relations~\citep{Thomas_2019_BH_galaxy_relation}; the group X-ray scaling relations~\citep{Robson_2020_X_ray_scaling_relation}; galaxy sizes and star formation rate profiles \citep{Appleby_2020}; damped Ly$\alpha$ abundances \citep{Hassan_2020}; the reionisation-epoch UV luminosity function \citep{Wu_2020}; the temperature--density relationship of the intergalactic medium \citep{Sorini_2020}; the galaxy colour bimodality in the stellar--halo-mass relation \citep{Cui2021} and the low-redshift Ly$\alpha$ absorption \citep{Christiansen_2020}.

We will consider two different variants from among the \simba-50 runs: `allphys' which includes all the physics above identical to the main \simba\ volume, and `nojet' which turns off both AGN jet and X-ray feedback.  The \simba\ suite includes other feedback variant models, including `no-feedback' where all feedback including galactic winds are off, `noagn' where only SF winds are included, and `no-X' where only X-ray AGN feedback is turned off but radiative and jet mode AGN feedback are on.  We examined these variants as well, but found that the only notable impact on the tSZ properties occurred when turning on/off the AGN jet feedback. This is consistent with previous results from \simba\ showing that the hot gas distribution in and around massive `allphys' haloes is by far the most sensitive to AGN jet feedback \citep{Christiansen_2020,Sorini_2021}. Thus for brevity, we eschew presenting the results for the other models here, and focus on only two feedback models from \simba-50: `allphys' and `nojet'.

We will primarily use the snapshot 141 from each \simba\ run, which spans a redshift range of $z=0.174-0.21$ (based on the redshift depth of the $100\hMpc$ volume) for \simba-100, and $z=0.174-0.192$ for the \simba-50 runs.  This is a good match to the redshift range spanned by the massive galaxies targeted by {\it Planck} for tSZ stacking.

\subsection{The TNG project}\label{ssec::tng_sim}

We also consider simulations from the TNG project\footnote{\url{https://www.tng-project.org/}} \citep{tng_sim_ref1, tng_sim_ref2, Nelson_2018_tng, tng_sim_ref3,tng_sim_ref4, tng_data_release, tng_galaxy_formation, tng_AGN_feedback}. This is a suite of magnetohydrodynamic simulations carried out with the $\textsc{Arepo}$ code \citep{arepo_ref}. Compared to the original Illustris project \citep{illustris_sim_paper}, it includes the magnetic fields \citep{tng_magneto_field_ref}, an improved version of the galaxy formation physics model \citep{tng_galaxy_formation,tng_sim_ref1} and an updated AGN feedback model \citep{tng_AGN_feedback}, and it was also calibrated to reproduce the observed galaxy stellar mass function~\citep{tng_galaxy_formation,tng_sim_ref1,Nelson_2018_tng}. Particles are evolved within a wide range of volumes and resolutions: boxes with volumes of $(35\hMpc)^{3}$ (denoted as TNG50-1, TNG50-2 and TNG50-3: ranked from the highest resolution to the lowest), $(75\hMpc)^{3}$ (TNG100-1, TNG100-2 and TNG100-3) and $(205\hMpc)^{3}$ (TNG300-1, TNG300-2 and TNG300-3). Dark-matter (hereafter DM)-only runs are also included as counterparts for these suites of simulations. The chosen cosmology for these simulations is the $\textit{Planck}$ 2015 cosmology \citep{planck_2015_cosmo_paper}, which is very close to the one assumed in the $\simba$ simulation. Here we use the TNG100 run with intermediate resolution (TNG100-2) for further analysis because the DM and gas particle resolution ($m_{\rm DM} = 4.0 \times10^{7} h^{-1}M_{\odot}$) are closest to those used in the $\simba$ simulations (see Table \ref{tab:sim_info}).

The adopted AGN feedback model in the TNG project \citep{tng_AGN_feedback} is a two-mode kinetic and thermal feedback model, which is similar to the one employed in the $\simba$ simulations. However, there are some major differences between those two. Instead of using the torque-limited accretion model for cold gas, TNG uses Bondi accretion for all phases. For modelling AGN feedback, at high Eddington rates, the gas surrounding the BH region is heated via spherical thermal feedback (as opposed to kinetic), and at low Eddington rates, the direction of the kinetic jet feedback is randomised at every timestep (as opposed to parallel/anti-parallel to the the angular momentum vector of the 256 closest gas particles); these choices tend to sphericalise the AGN feedback energy input more than $\simba$'s implementation. Despite these differences, the feedback models adopted by both simulations successfully reproduce a similarly broad range of observations over cosmic time.

\subsection{Construction of the thermal SZ \textbf{\textit{y}}-map}\label{sec::reconstrunction_of_y_map}

\begin{figure*}
\hspace{-4 cm}
  \begin{minipage}[b]{0.99\textwidth}
        \centering
        \includegraphics[width=1.2\linewidth]{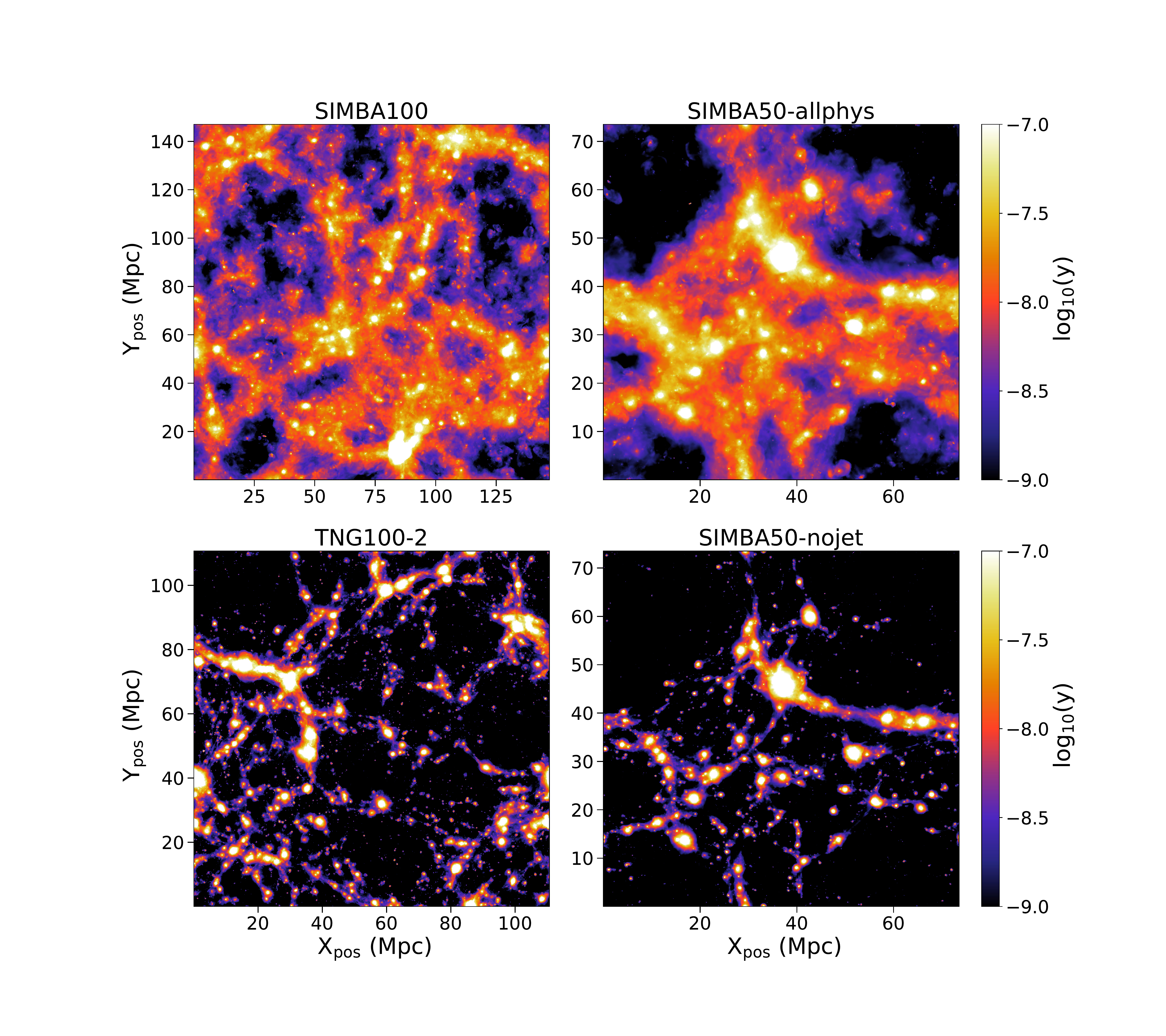}
        \end{minipage}
        \vspace{-1 cm}
\caption{2D tSZ-$y$-map constructed from the $\simba$100 model (\textit{upper left}), the $\simba$50 `allphys' model (\textit{upper right}) which includes a jet feedback model, the TNG100-2 simulation (\textit{lower left}) and the $\simba$50 `nojet' model (\textit{lower right}) at $z = 0.17$. Each pixel value of $y$ derives from a line-of-sight integration with a $50\hMpc$ total depth around the box centre. Map resolutions are $10 \,h^{-1} \rm ckpc$ for $\simba$ and $15\,h^{-1} \rm ckpc$ for TNG100-2. The colourbar shows the $y$-parameter in $\rm log_{10}$ scale. The impact of different energetic feedback models on the gas distribution in the IGM is noticeable.}
\label{stacked_SZ_map_along_zdir}
\end{figure*}

From the gas element data in the simulations, we construct simulated tSZ $y$ maps, and use these to generate synthetic observables that will allow us to make comparisons with data and set constraints on galaxy formation  models.

Projected $y$-maps are computed by using the \texttt{PyMSZ} package\footnote{\url{https://github.com/weiguangcui/pymsz}} \citep{pymsz_ref_paper}. Based on Equation \ref{eqn:y_params_eq}, the $y$-parameter is computed by the integral of the electron thermal pressure along the line of sight, which in the simulations corresponds to the summation of all the gas particle cells along the entire box. Specifically, the integral in Equation~\ref{eqn:y_params_eq} is discretized as follows \citep{pymsz_package_ref_supp,pymsz_ref_paper}:
\begin{equation}\label{eqn:y_params_cal_in_sim}
   y = \frac{\sigma_{\rm T}k_{\rm B}}{m_{e}c^{2}dA}\sum_{i} T_{i}N_{e,i}W(r, h_{i}), 
\end{equation}
where $N_{e,i}$ is the number of electrons per gas particle cell given by $N_{e,i}  = n_{e,i}dAd\ell$, where $n_e$ is the electron number density and $dA, d\ell$ are the chosen projected pixel area and line-of-sight distance unit. $W(r, h_{i})$ is the cubic spline kernel enclosing 64 non star-forming gas elements (as used in \simba), accounting for smoothing the tSZ $y$ signal onto the image pixels.  In simulations,
\begin{equation}
N_{e,i} = \frac{N_{e, H}\,m_{i}\,(1-Z-Y_{\rm He})}{\mu m_{\rm p}} \, ,
\end{equation}
where $N_{e, H}$ is the electron abundance per gas particle defined as the fractional electron number relative to the total hydrogen number, $\mu$ is the mean molecular weight, $m_{\rm p}$ is the mass of proton, $Z$ is the metal mass fraction of the gas particle and $Y_{\rm He}$ is the helium mass fraction; these values are all tracked directly in these simulations. The temperature of gas particles $T_{i}$ is computed from the specific thermal energy $U$ via 
\begin{equation}\label{eqn::temp_cal}
T_{i} = (\gamma-1)\frac{U m_{p}\mu}{k_{\rm B}}, 
\end{equation}
where $\gamma=5/3$ is the adiabatic index for monoatomic gases.

Figure \ref{stacked_SZ_map_along_zdir} shows the 2D $y$-maps of different models derived from a line-of-sight integration with a $50\hMpc$ total depth around the box centre at $z = 0.17$. The top row shows the primary $100\hMpc$ volume on the left, and the $50\hMpc$ one with the identical input physics ($\simba$50 `allphys') on the right. The two boxes of simulations with the same input physics will be used for a volume convergence test of our results. The lower left panel shows the $y$ map from the TNG100-2 run.  The volume is intermediate between $\simba$50 and $\simba$100, but there is clearly less widespread IGM heating in this model. Finally, the lower right panel shows the $\simba$50 `nojet' results with AGN jet feedback (and associated X-ray feedback) turned off. In $\simba$, this dramatically reduces the IGM pressure~\citep{Christiansen_2020}. The TNG100-2 map looks qualitatively more similar to the $\simba$50 `nojet' map.

It is evident that the black hole jet model in $\simba$ `allphys' is efficient in pumping up the pressure of the gas, and distending it beyond the haloes into the IGM. This gives $\simba$ `allphys' the most extended distribution of gas pressure, followed by TNG100-2 and then $\simba$50 `nojet'. This result is not surprising because the AGN model adopted in TNG simulation is less efficient at expelling gas into the IGM compared to that used by $\simba$, which causes an overprediction of the hot gas fraction in TNG groups and clusters \citep[as discussed in e.g.][]{Barnes_2018_tng_hot_gas,simulation_review_hgf}. This is consistent with the resulting tSZ maps shown in Figure \ref{stacked_SZ_map_along_zdir}. These differences in the simulated tSZ maps across different models suggest great potential to distinguish feedback models using $y$-parameter statistics, which will be the main focus of our next section.

\section{The \textbf{\textit{Y}} -- \textbf{\textit{M}} relation}\label{sec::Y_M_relation}

\begin{figure*}
    \begin{minipage}[b]{0.99\textwidth}
        \centering
        \includegraphics[width=0.99\linewidth]{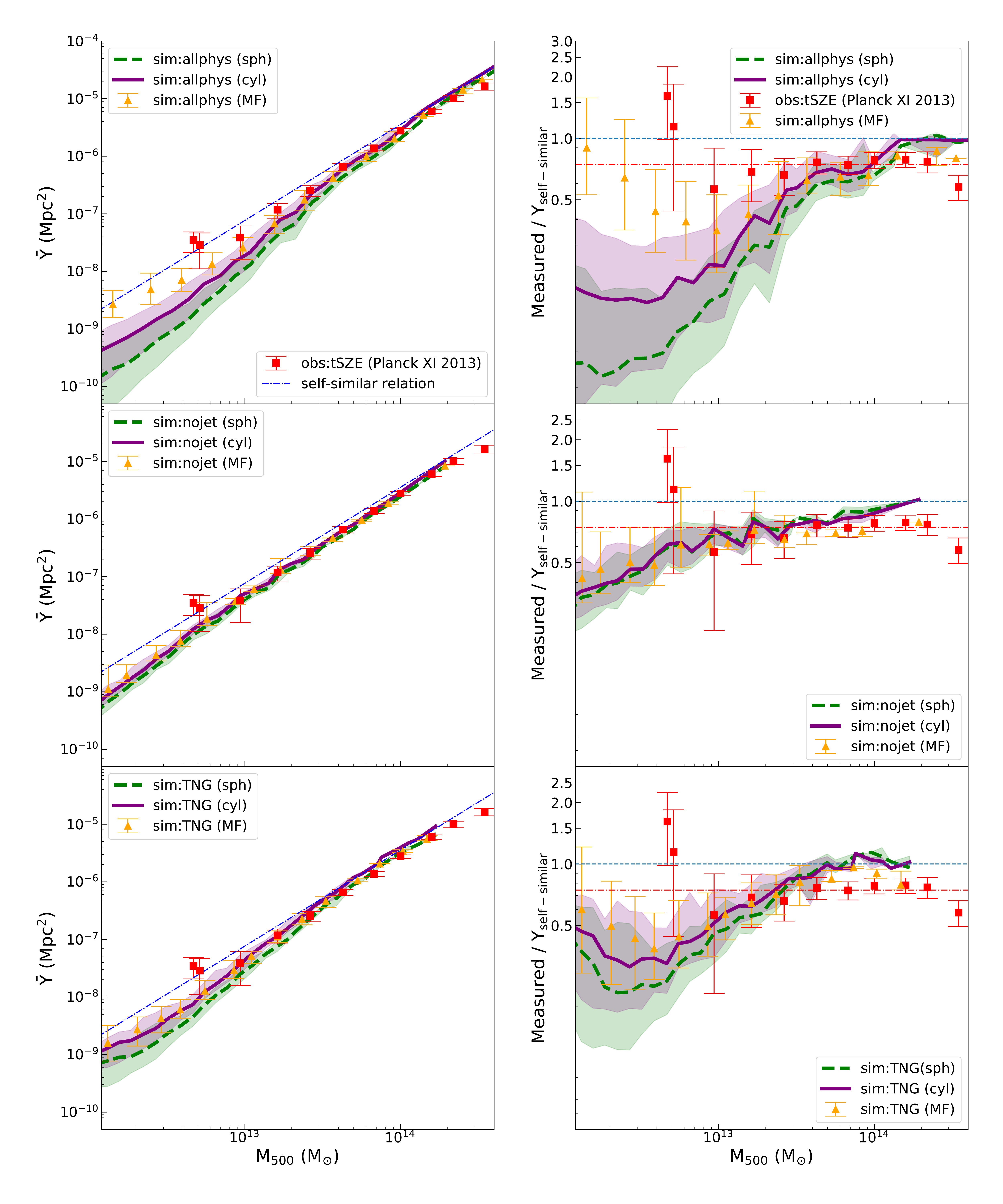}
    \end{minipage}%
    \vskip-0.2in
\caption{\textit{Left}: comparison of the measured $Y$--$M_{500}$ relation between simulations (${Y_{\rm sph}}$, green dashed: averaging within a sphere; ${Y_{\rm cyl}}$, purple solid: projection along a cylinder; orange data points with errors: measurement using the Matched Filter technique on the simulated $y$-map. See the main text for more details) and observations (red squares with errors: \citet{Planck_Y_M_paper}; blue dot-dashed: best-fit self-similar model from X-ray observations of \citet{Arnaud_paper}). \textit{Right}: ratio between the measured $Y$--$M$ relation and the self-similar model. For comparison, the ratio between the best-fit scaling relations of \citet{Planck_Y_M_paper} and \citet{Arnaud_paper} is shown as red dot-dashed line. The shaded regions are 1$\sigma$ Poisson errors from simulations.}
\label{Y_M_relation_halo_mass}
\end{figure*}

\begin{figure*}
 \hspace{-3.5 cm}
    \begin{minipage}[b]{1.0\textwidth}
        \centering
        \includegraphics[width=1.2\linewidth]{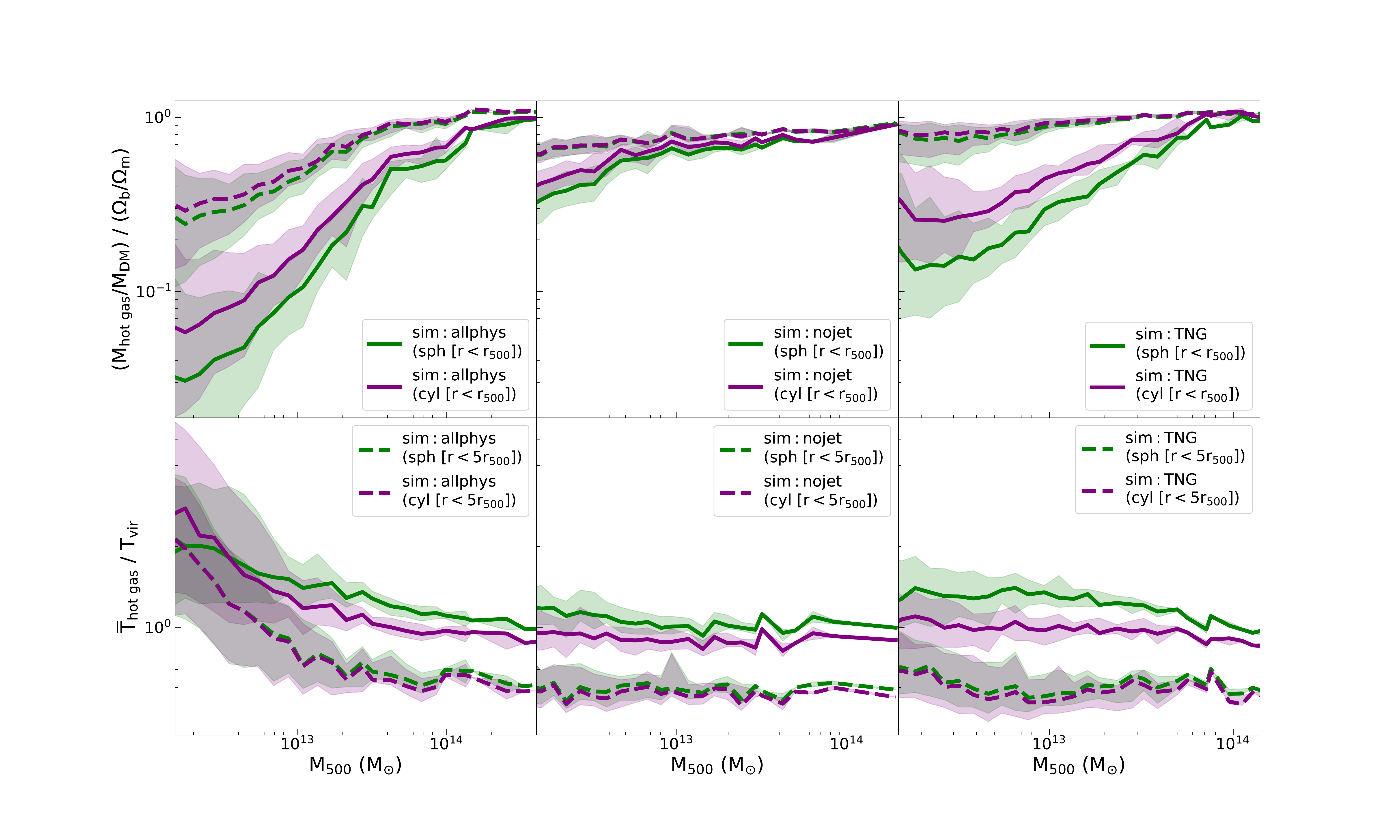}
    \end{minipage}%
    \vskip-0.3in
\caption{Hot gas fraction (\textit{upper}) and normalised mean temperature (\textit{lower}) around haloes computed within a sphere with radius $R = r_{500}$ (green solid line) and within a cylinder with radius $R = r_{500}$ and a length of plus-minus $5r_{500}$ (purple solid line). As indicated in the legends, dotted lines are for a larger radius $5 r_{500}$. Hot gas are gas particles with temperatures greater than $10^{5}$K, where the temperature value per gas cell is computed using Equation \ref{eqn::temp_cal}. The error bars for the simulations indicate the 68 percentile ranges of the median per halo mass bin.}
\label{hgf_T_comp}
\end{figure*}

Using current observations such as from {\it Planck\/}, the $Y$--$M$ relation around haloes has been measured via stacking down to $M_{500}\sim 5\times10^{12}M_{\odot}$. However, as mentioned in the introduction, a major challenge in using these observations to constrain galaxy formation models and the baryonic content is to disentangle the effect of gas physics from theoretical and observational systematics. We focus on the projection effect from large-scale structure and on the convolution of the beam in CMB observations. With the listed simulations, we approach this with the following three steps:
\begin{enumerate}
    \item We establish the `ground truth' predictions for the $Y$--$M$ relation from simulations by measuring $Y_{\rm sph}(R)$, the total tSZ signal integrated within a radius $R$, where $R=r_{500}$ is chosen. This is the most idealised situation where $Y_{\rm sph}(R)$ is contributed by the gas within the halo, and no projection effect from neighbouring halo or large-scale structure contribute. Note that the $Y_{\rm sph}(R)$ is calculated by direct summation of $y$ within $r_{500}$ \citep[see also Equation 14 in][]{Arnaud_paper}. \\
    \item To account for the effect of line-of-sight projection either from neighbouring haloes or from gas in the large-scale structure, we measure the cylindrically integrated $Y_{\rm cyl}(R)$ using our simulated $y$-maps. This characterises the tSZ flux within a aperture $R$, where $R = r_{500}$, and $R_{b} = 5r_{500}$, where $R_{b}$ is the depth of projection along the line-of-sight from the halo \citep{Arnaud_paper}. It is therefore expected that $Y_{\rm cyl}(R)>  Y_{\rm sph}(R)$ by design.\\
    \item To investigate the effect of finite resolution in observations, we convolve our simulated $y$-maps with Gaussian beams to mimic observational data, and perform similar analyses as in observations, such as the matched filters (MF). This will allow us to make direct comparison with observations to test the models. This will be detailed in Section~\ref{sec:obs}.
\end{enumerate}

Figure~\ref{Y_M_relation_halo_mass} shows these comparisons for $M_{\rm 500}>10^{12.3}M_\odot$ at $z = 0.17$, and this encapsulates the central results of our paper.  For comparison, the redshift of the simulation output is chosen to be approximately the same as the median redshift of the bright galaxy sample presented in \citet{Planck_Y_M_paper}.

Here we show the results for $Y_{\rm sph}(R)$ as the green dashed line with $1\sigma$ shading, $Y_{\rm cyl}(R)$ as the purple solid line with purple shading, and $Y_{\rm MF}$ as the orange data points with error bars. For $Y_{\rm sph}(R)$ and $Y_{\rm cyl}(R)$, we integrate over $r_{500}$ for haloes down to the lowest masses probed here. $Y_{\rm MF}$ is calculated by fully mimicking the {\it Planck\/} observations: a beam size of $10'$, an aperture size of $5 r_{500}$ and the MF analysis that we describe below. These orange data points may thus be directly compared to the {\it Planck\/} stacked tSZ measurements shown as the red data points.  The blue dot-dashed line shows the self-similar scaling extrapolated from \citet{Arnaud_paper}.

From the top to bottom rows, the left panels show the $Y$--$M_{500}$ relation for the full \simba-100 run, the \simba-50 `nojet' run, and TNG100-2, respectively.  In the right panels, we highlight the relative differences by (arbitrarily) normalising to the \citet{Arnaud_paper} self-similar relation (blue dotted line at unity).  We also show the best-fit self-similar scaling of the \citet{Planck_Y_M_paper} data as a red dot-dashed line.  We do not show the \simba-50 `allphys' run since it is generally similar to the \simba-100 run; we explore numerical convergence for \simba\ and IllustrisTNG in Appendix~\ref{sec::convergence_test_full_physics_models}.

Figure~\ref{Y_M_relation_halo_mass} shows that for $Y_{\rm sph}(R)$ at the high mass end, all simulation results converge to the self-similar model (blue line). This indicates that the impact of feedback on the gas around massive haloes with $M_{500} \ga 10^{14} M_{\odot}$ is relatively minor. Towards the low-mass end, models with strong feedback show increasing deviations from the self-similar model, such as the $\simba$ full physics model in the top panel and TNG100-2 in the bottom. This can be understood as the relatively strong influence of the AGN feedback for lower mass haloes, where the gravitational potentials are shallower and they tend to lose their gas more easily owing to the kinematic energy injected from AGN feedback. The \simba full physics model predicts stronger deviations from the self-similar model than TNG does, as seen more clearly in the right-hand panels of Figure~\ref{Y_M_relation_halo_mass}. These results are also consistent with those reported from \cite{Le_Brun_paper} with strong AGN feedback models. 

A comparison between $Y_{\rm cyl}(R)$ and $Y_{\rm sph}(R)$ provides insights into the $y$ contribution from the gas distribution in surrounding haloes and the IGM. $Y_{\rm cyl}(R)$ (purple line) is consistently larger than $Y_{\rm sph}(R)$ (green line) across different models. This is because $Y_{\rm cyl}(R)$ takes contributions from the outer regions of haloes while $Y_{\rm sph}(R)$ does not. For the \simba-100 and TNG100-2 models, the difference between $Y_{\rm cyl}(R)$ and $Y_{\rm sph}(R)$ increases with decreasing halo mass. This suggests that the contribution from gas outside the halo becomes increasingly strong for lower mass haloes. In contrast, the \simba-50 `nojet' model predicts very similar $Y_{\rm cyl}(R)$ and $Y_{\rm sph}(R)$, and they both have similar slopes to the self-similar model. This indicates that without jet feedback in \simba (but still with radiative AGN feedback and star formation feedback included), the hot gas predominantly resides within the main haloes across all different masses in this model, and that there is a very minor injection of non-thermal energy into the gas, or evacuation of the hot halo gas. In Section \ref{sec:origin}, we will quantify the physical conditions in and around haloes to investigate the origin of the difference. 

When applying the same processes as {\it Planck\/} to the mock maps, we see the $Y_{\rm MF}$--$M_{500}$ relation from all three simulations (orange data points with error bars) follows the {\it Planck\/} result well. There is no obvious distinction between the predictions of the three models. This was also seen in \citet{Le_Brun_paper}, where they performed the same MF method as {\it Planck\/} to recover the $Y_{\rm MF}$--$M_{500}$ relation on tSZ maps from different cosmo-OWLS models. All of their models appear to be similar to the self-similar model, which means that results from their AGN models are biased against `the truth' ($Y_{\rm sph}$--$M_{500}$ ). They attribute this to the use of an inappropriate gas pressure profile. We will discuss this in more detail in Section \ref{sec:obs}. In this study, we will focus on the effects of beam smoothing. In the same section (Section \ref{sec:obs}), we describe our MF procedure for creating mock {\it Planck\/} data, we investigate the effects of beam smearing, and we discuss its impact on interpreting the baryonic content of massive haloes.
 
\subsection{The origin of deviations from self-similarity} \label{sec:origin}

To understand the origin of the differences between the predicted $Y$ signal from various models versus self-similar expectations, we examine the two key quantities that go into computing $Y$: the gas density and gas temperature within and around massive haloes.  In the self-similar case, one assumes that halo gas is heated purely via gravitational shocks. Deviations from self-similarity in either temperature or density reflect non-gravitational processes that will affect the tSZ decrements.

Figure \ref{hgf_T_comp} shows the hot baryon fraction scaled to the cosmic baryon fraction (top row) and the mean hot gas temperature (bottom row) normalised by the virial temperature as a function of $M_{500}$. We assume that the virial temperature versus halo mass follows a self-similar relation with a slope of 2/3 \citep{kaiser_cluster}, and the curve is normalised to the most massive systems. The three columns correspond to the $\simba$100, $\simba$50-`nojet', and TNG100-2 models, respectively (we have confirmed that $\simba$100 and $\simba$50-`allphys' are similar).  The green solid and dashed lines show the values computed within a spherical aperture of $r_{500}$ and $5r_{500}$, respectively, and the purple lines correspondingly show the cylindrical aperture at those radii, within plus-minus $5r_{500}$ along the line of sight.  The shading shows the $16-84\%$ ranges around the median for each halo mass bin. 

Looking at the hot gas fractions, it is clear that they are dramatically affected by AGN jet feedback.  At cluster masses, haloes contain roughly their cosmic fraction of baryons in hot gas.  But below $M_{500}\la 10^{14} M_\odot$, the hot gas fraction becomes increasingly suppressed with full $\simba$ physics. \citet{simba_ref} and \citet{Appleby_2021} showed that this hot gas deficit is not accounted for by cold gas and stars, and instead reflects a genuine deficit in baryon fractions in these haloes. At $M_{500}\sim 10^{12.5}M_\odot$, haloes contain a hot baryon fraction of $<10$\% within $r_{500}$ in \simba-100, but $\sim 50\%$ in \simba-50 `nojet'.  TNG100-2 is intermediate between these. Moving to a larger radius, $5r_{500}$, which are more comparable to the {\it Planck\/} beam, the baryon fractions are less different at low masses, but the full $\simba$100 run still shows a substantial deficit.  This indicates that \simba's jet feedback is able to push out baryons even beyond $5r_{500}$, as shown by \cite{Sorini_2021}. Meanwhile, at these larger radii, TNG100-2 looks more similar to \simba-50 `nojet', indicating that TNG's feedback model does not have such a widespread effect.  Nonetheless, the models look much more similar at $5r_{500}$ (dashed lines) than at $r_{500}$, indicating that observational discrimination between the models would be more optimal at higher resolution than typically provided by {\it Planck\/}. Due to the limited resolution of {\it Planck\/}, the tSZ flux is measured within a much large aperture size of $5r_{500}$, and $Y_{5r_{500}}$ is converted to $Y_{r_{500}}$. This relies on the assumption of the spatial profile from \citet{Arnaud_paper} \citep[see][]{Planck_5r500_r500_convert}, and thus may not be robust \citep{Le_Brun_paper}. We will investigate this point further in the next subsection.

At low masses ($M_{500} \la 10^{13} M_{\odot}$) the error bars become large owing to the wide range of hot gas contents in these haloes, so robust discrimination between models may become more difficult here.  Thus in $\simba$, the impact of AGN feedback might be the most evident at $M_{500} \sim 10^{13-14} M_{\odot}$, suggesting that haloes in this mass range would be ideal candidates for distinguishing between feedback models. This is consistent with the results in \citet{Lim_tng_eagle_mag_sim_paper}, that the measured gas profiles across simulations are the most different for haloes with $M_{500} \sim 10^{13-13.5} M_{\odot}$.

The temperature shows less dramatic differences between the various models (lower panels of Figure~\ref{hgf_T_comp}). In general, \simba-50 `nojet' and TNG100-2 show less deviation from the self-similar assumption, which is expected due to their relatively weak feedback effects. \simba-100 shows an increasing deviation towards low halo mass, highlighting the impact of AGN jet heating. This again indicates more widespread impact of feedback in this model. For the \simba-100 model, the gas fraction is $\sim\,$0.1 in the low halo mass bins. However, from Figure \ref{Y_M_relation_halo_mass}, the $Y$--$M$ relation deviates from the self-similar prediction by a factor of $\sim\,$0.3. This increase in temperature compensates somewhat for the strong hot gas deficit at the low halo mass, which shifts the $Y$--$M$ relation a little upwards.

Juxtaposing the $Y$--$M$ relation in Figure~\ref{Y_M_relation_halo_mass} versus these plots, it is clear that the departures from self-similarity in $Y$--$M$ are primarily driven by the hot baryon fractions within haloes rather than changes in temperature.  The suppression in $Y$ at low $M_{500}$ relative to self-similarity is qualitatively similar to the suppression of the hot baryon fraction, though quantitatively it is somewhat less because it is mildly countered by the increased temperature at lower halo mass, especially when using the $5r_{500}$ aperture. Similar results have been reported by several previous studies, even for massive clusters. \citet{Barnes_2017} studied several scaling relations of massive clusters (with $M_{500}\ga10^{14}M_{\odot} $) from simulations. They found that the best-fit mass slope of the $Y_{\rm SZ}$--$M_{500}$ relation is 13\% steeper than the self-similar prediction at $z = 0$, which is mainly driven by the gas expulsion due to AGN activity in clusters. \citet{Le_brun_2017} also studied the evolution of the $Y_{\rm SZ}$--$M_{500}$ relation for clusters (with $M_{500}\ga10^{13}M_{\odot} $) using different cosmo-OWLS model variants. Similarly, they found that efficient AGN feedback is the main factor accounting for the significant deviation from self-similarity, and that this diminishes for more massive clusters or when feedback activities are excluded. They also found that the deviation of the recovered $T$--$M_{500}$ relation from the self-similar prediction is much smaller than the deviation of the $M_{\rm gas}$--$M_{500}$ relation, which is consistent with \citet{Barnes_2017} and our findings here. Similar findings are also discussed in other studies \citep[e.g.][]{Fable_Y_x_M500}. Thus we confirm that the $Y$--$M$ relation is a sensitive probe of the baryon fraction in haloes at $M_{500}\la 10^{14}M_\odot$, which can be constrained by future tSZ surveys, but one must account for any feedback heating associated with gas expulsion in order to interpret this properly.

In summary, both the $Y_{\rm sph}$--$M$ and $Y_{\rm cyl}$--$M$ relations from simulations with strong AGN feedback predict deviations from the self-similar model. Our results are consistent with other simulations such as cosmo-OWLS-AGN, in which the effects of strong feedback were also examined \citep{Le_Brun_paper}. At face value, this seems to be at odds with the measurements from {\it Planck\/}, which are consistent with self-similarity at all masses. However, for fair comparison with observations, it is important to apply the same analysis procedure. We will focus on this in the next subsection.

\begin{figure*}
\hspace{-1. cm}
    \begin{minipage}[b]{1.0\textwidth}
        \centering
        \includegraphics[width=1.05\linewidth]{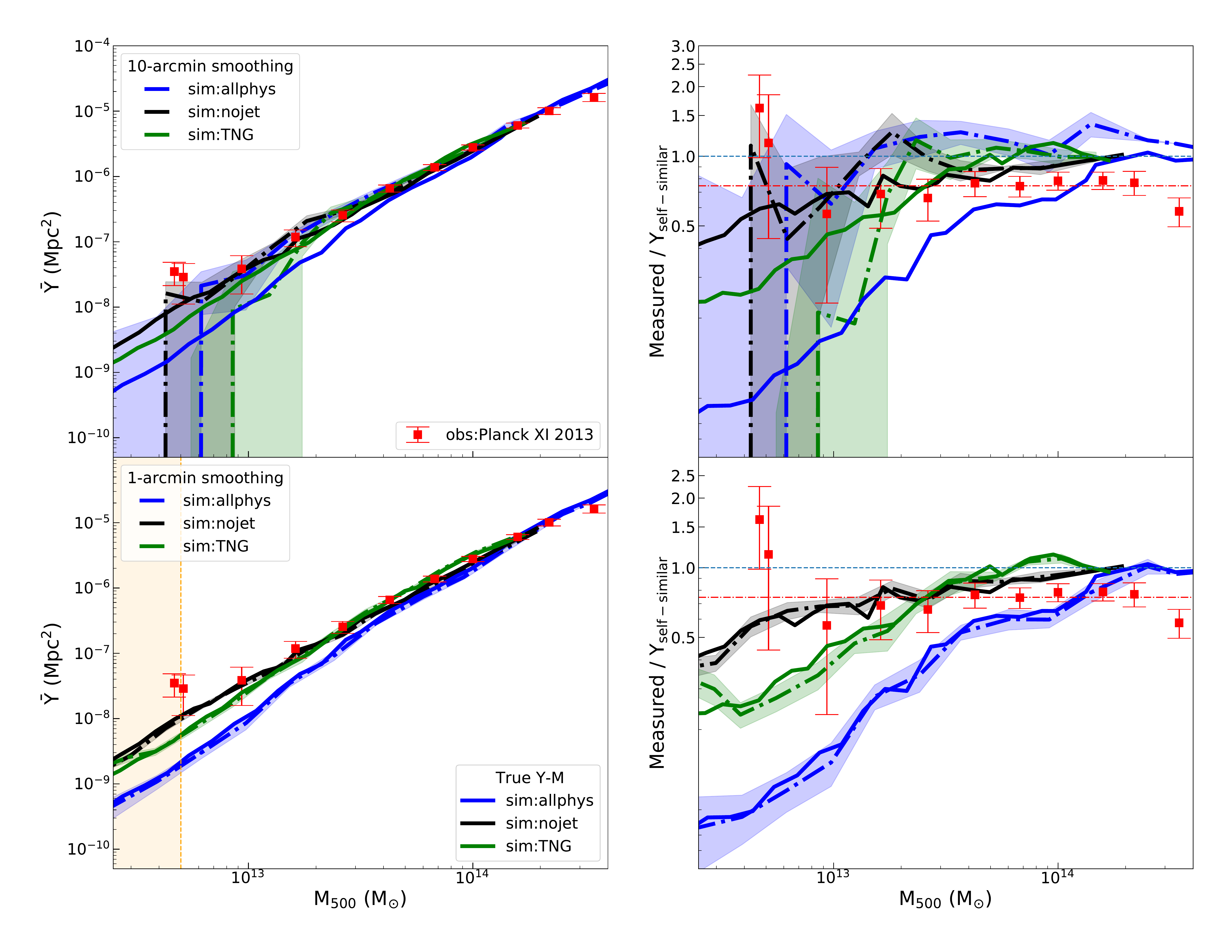}
    \end{minipage}%
    \vskip-0.2in
\caption{Comparison of the measured $Y$--$M$ relations using the aperture photometry (AP) method on the simulated maps (see text for details). Solid lines show the expected \textit{true} $Y$--$M$ curves measured within apertures of ($r_{500}$, $\sqrt{2}r_{500}$) on unsmoothed maps, while the dot-dashed lines show the $Y$--$M$ curves measured on maps smoothed by 2D Gaussian beam with FWHM = 10 arcmin (upper) or FWHM = 1 arcmin (lower). Aperture sizes for these two cases depend on whether the inner $r_{500}$ region of haloes can be resolved by the chosen beam size. Orange vertical line on the lower panel indicates the lowest bound that a halo can be resolved by a 1-arcmin beam, while no haloes can be resolved by a 10-arcmin beam within the considered mass range. Different models are distinguished by colours, and the shaded area shows the 68 percentile range around the measured median. For comparison, the ratio between the measured $Y$--$M$ curves and the self-similar model is shown on the right panel.}
\label{beam_size_comp}
\end{figure*}

\subsection{The effect of beam smearing}
\label{sec:obs}
The observational results of the $Y$--$M$ relation (red data points with errors) presented in \citet{Planck_Y_M_paper} come from applying a Matched Filter technique (MF) to the reconstructed tSZ $y$-map \citep{Melin_MF_ref}. In Fourier space, the filter function is 
\begin{equation}\label{eqn::MMF_filter}
    \hat{F}(\textbf{\textit{k}}) = \bigg[\int \frac{|\hat{\tau}(\textbf{\textit{k}}')\hat{B}(\textbf{\textit{k}}')|^{2}}{P(k')} \frac{d^{2}k'}{(2\pi)^2} \bigg]^{-1} \frac{\hat{\tau}(\textbf{\textit{k}})\hat{B}(\textbf{\textit{k}})}{P(k)},
\end{equation}
where $\hat{\tau}(\textbf{\textit{k}})$ is the Fourier transform of the pressure profile of the halo; $\hat{B}(\textbf{\textit{k}})$ is the window function that mimics the $\textit{Planck}$ beam \citep{beam_func_ref}. $P(k)$ is the noise power spectrum of the $\textit{Planck}$ NILC $y$-map. Therefore, the filtered $Y$ signal is affected by all three ingredients of the filter. With stacking, we expect the noise to be sub-dominant.

We then adopt the same ingredients as in \citet{Planck_Y_M_paper} for Equation~\ref{eqn::MMF_filter}, apply the MF technique to our simulated $Y_{\rm cyl}(R)$ maps, and use it to extract a mock-{\it Planck\/} $Y$ signal around the haloes. Specifically, we employ the universal pressure profile derived from X-ray observations and hydrodynamic simulations \citep{Arnaud_paper}, the $10'$ FWHM Gaussian beam and the noise spectrum of the $y$-map from {\it Planck\/}. This results in the orange data points with error bars shown in Figure~\ref{Y_M_relation_halo_mass}.

This plot demonstrates that all the deviations from the self-similar model essentially disappear after the MF filtering. The filtered signals for all different models are broadly consistent with each other and with the observations from {\it Planck\/}. This result is consistent with that of \citet[see their Figure 2]{Le_Brun_paper}, where MF was applied to their simulated $y$-maps. It appears that the filtering process has washed away the variations between different models observed in the previous section. The question is, which ingredients of the filter are the main cause?

\citet{Le_Brun_paper} pointed out that the assumed universal pressure profile \citep{Arnaud_paper} may not represent the true pressure profiles of the data, and it is certainly different from the pressure profiles seen in their simulations. After adopting their simulation pressure profiles, their predicted $Y$--$M$ relation from the fiducial AGN model (AGN8.0) is in approximate agreement with the {\it Planck\/} data \citep[See Figure 4 in ][]{Le_Brun_paper}. Also, the beam used to convolve their simulated $y$-maps was kept fixed. Given that the size of the beam for the {\it Planck\/} $y$-map is 10\,arcmin\footnote{The beam of the {\it Planck\/} tSZ $y$-map is ${\rm FWHM}=10\,$arcmin, as stated in \citet{planck_tsz_paper}; however in \citet{Lim_Y_M_paper}, a 5-arcmin Gaussian beam was adopted instead.}, which is larger than the $r_{500}$ of most haloes, we suspect that the exact form of the pressure profile $\hat{\tau}(\textbf{\textit{k}})$ may not strongly affect the shape of the $Y$--$M$ relation. Indeed, as also noted in  \citet{Planck_Y_M_paper,AP_ref}, even if haloes are taken as point sources, the resulting slope of the $Y$--$M$ relation from the filtering is `practically unaffected' (but the normalisation will change slightly). Perhaps for the same reason, when a compensated top-hat filter is used (or aperture photometry, AP, as noted in \citealt{Planck_Y_M_paper}) instead of the MF filtering, the recovered $Y$--$M$ relation remains very similar to the one obtained using their MF \citep{Planck_Y_M_paper}.

Specifically, the AP method proposes that, for a halo with radius $R$, its integrated $Y$ can be estimated by subtracting the total tSZ flux within an aperture with size $R$ from the flux within a ring of inner radius $R$ and outer radius $fR$, where the latter is used to account for the contribution of the background. When $f=\sqrt{2}$, the background and the signal are estimated within areas of the same size. In this way, the residual within the inner radius $R$ after the subtraction of the background is approximately the signal generated by the halo itself. We can see that this method depends only on the mean fluxes of two broad regions around the halo, and should be insensitive to the gas pressure profile of the halo.

In light of the above, we adopt a similar aperture filtering process as in \citet{Planck_Y_M_paper} to reduce the sensitivity to the exact shape of the pressure profile. This allows us to focus on testing the impact of the beam. We convolve our simulated $y$-maps at $z = 0.17$ with a Gaussian beam of ${\rm FWHM}=10'$ and $1'$ to approximate the beam of {\it Planck\/} and the South Pole Telescope (SPT) or the Atacama Cosmology Telescope (ACT) \citep{SPT_survey_ref,act_ref} respectively. 

Figure~\ref{beam_size_comp} shows the results of this comparison of beam sizes using the AP method.  For the case of ${\rm FWHM}=10'$, $r_{500}$ regions are mostly unresolved, especially for low-mass haloes. AP signal remains similar to the cases of MF, and is consistent with the self-similar model (upper panel of Figure~\ref{beam_size_comp}). This suggests that despite the drastic impact of feedback in some of the models, the distributions of gas pressure around haloes are indistinguishable after smoothing with ${\rm FWHM}=10'$. 

For ${\rm FWHM}=1'$ however, haloes with 
$M_{500}>5\times 10^{12}M_{\odot}$ can be resolved. The filtered $Y$--$M_{500}$ relations are then in excellent agreement with the true $Y$--$M_{500}$ relations for all three models, and all three models can be distinguished from each other at the low mass end (lower panel of Figure~\ref{beam_size_comp}). We have also repeated the analysis with the two different smoothing scales (${\rm FWHM}=1'$ and $10'$), but using the MF technique. The results are very similar to those with the AP technique. Therefore, it is evident that having high resolution for the CMB maps will allow us to recover the near-ground truth predicted $Y$ values from simulations, and this is the key to obtaining constraints on models of galaxy formation from the $Y$--$M_{500}$ relation.

In summary, we have observed that state-of-the-art hydro-simulations, such as \simba and TNG, predict a deviation from the self-similar relation for low-mass haloes, but only if the true $Y$ signal within the radii of the haloes or a comparable fixed aperture can be measured. However, once convolved with a 10-arcmin Gaussian beam as in the {\it Planck\/} $y$-maps, the differences in the $Y$--$M_{500}$ relation among different models are washed away. All models seem to predict similar $Y$--$M_{500}$ relations, and they appear to be consistent with the self-similar model, as well as consistent with the observation from {\it Planck\/}. When a finer Gaussian beam such as 1\,arcmin was used instead, the filtered $Y$--$M_{500}$ relation can successfully recover the ground truth predictions from our simulations, and different models of galaxy formation can be discriminated.

\section{Discussions and Conclusions}\label{sec::conclusion}
AGN feedback is one of the most uncertain aspects of galaxy formation models, with qualitatively different models yielding similar predictions for galaxies. In this paper, we have investigated how different feedback models affect the distribution of baryons within haloes, by studying the measured integrated thermal Sunyaev Zeldovich $y$ decrement versus halo mass in cosmological hydrodynamic simulations, and comparing them with observations from {\it Planck\/}. We aim to understand the systematics in using tSZ observations, especially the $Y$--$M_{500}$ relation, to distinguish the different baryonic feedback models. Using the \simba and TNG100-2 simulation data, we found that at low mass ($M_{\rm 500}\la 10^{14}M_\odot$), the true $Y$--$M_{500}$ relations drops below the power-law self-similar prediction when energetic AGN feedback is included. In \simba, this deviation is mainly due to the halo baryon fractions being lowered by \simba's AGN jet feedback. This indicates that deviations from self-similarity in $Y$--$M$ can test feedback physics.  However, such comparisons are complicated by observational systematics such as line-of-sight projection and beam smearing, and in this work we have quantified these effects using simulations.

We summarize our key findings below:
\begin{itemize}
    \item The total tSZ decrement within a spherical aperture at $r_{500}$, $Y_{\rm sph}$, versus halo mass $M_{500}$ shows strong deviations from self-similarity for the full \simba model, less strong for TNG100-2, and essentially no deviation for a \simba run without AGN jet feedback. The deviation generally grows towards lower halo masses, indicating a stronger effect of AGN feedback in smaller systems.
    \item When including line-of-sight projection effects by considering cylindrical apertures at $r_{500}$, these trends are weakened slightly, but still evident.
    \item When we follow the analysis to mimic the {\it Planck\/} data for our simulated $y$-maps, we find that the self-similar relation is recovered for all the models in our investigation, and all are in good agreement with {\it Planck\/} data down to the lowest probed halo masses ($M_{\rm halo}\sim 10^{12.5}M_\odot$).
    \item We thus identify that the effective convolution of the {\it Planck\/} $y$-maps with a relatively large beam is the main reason for the agreement with {\it Planck\/},  although projection effects are not negligible particularly for lower-mass systems. 
    \item With a finer beam ($\sim\,$1\,arcmin) for the $y$-maps, the true $Y$--$M_{500}$ curves predicted from simulations are recovered with both the AP and MF methods. Thus, there are good grounds for expecting that future tSZ observations with a high angular resolution will be able to distinguish these different baryon models, especially at the group halo mass scale. 
    \item We further show that the deviations from self-similarity for $Y_{\rm sph}$ and $Y_{\rm cyl}$ are driven primarily by AGN feedback lowering the halo hot baryon content. Therefore future high-resolution measurements of the $Y$--$M$ relation at low masses will most directly constrain the hot baryon fractions in haloes.
\end{itemize}

Our findings suggest that one can constrain the hot halo gas fraction, and thereby differentiate between different AGN feedback models, from the low-mass end of the $Y$--$M_{500}$ relations measured with high resolution tSZ surveys.  Many forthcoming surveys with higher-resolution tSZ data could thus provide better constraints on the galactic feedback models. The Atacama Cosmology Telescope (ACT) and the South Pole Telescope (SPT) can achieve beams of FWHM $\approx 1$ arcmin \citep{act_ref,SPT_survey_ref}, and the NIKA2 can even achieve $\sim\,$15$''$ resolution \citep{NIKA2}, while future facilities such as AtLAST can reach $\sim\,$5$''$ in (sub-)millimetre bands \citep{AtLAST_ref}. Other CMB experiments such as the Simons Observatory \citep{SO_ref}, CCAT-prime \citep{CCAT_Prime_ref}, and CMB-S4 \citep{CMB_S4_ref} will conduct observations using more frequency bands, which is extremely helpful for separating a clean tSZ signal from other potential contaminants. Such high sensitivity data will allow us to probe lower-mass haloes, as well as resolving the inner regions of haloes. 

With such future higher-resolution tSZ measurements, when combining the $Y$--$M_{500}$ relation with other measures such as kinetic SZ, SZ two-point statistics, and SZ profiles, it will be possible to place even stronger constraints on the galactic feedback process. Hence, SZ observations provide a promising and exciting way to study the distribution of hot baryons in the Universe.

\section*{Acknowledgements}

We are grateful for the publicly available simulations from the \simba\ and IllustrisTNG projects. 
Throughout this work, DS and JAP were supported by the European Research Council, under grant no. 670193, and the STFC consolidated grant no. RA5496. 
WC is supported by the STFC AGP Grant ST/V000594/1. He further acknowledges the science research grants from the China Manned Space Project with NO. CMS-CSST-2021-A01 and CMS-CSST-2021-B01.
RD acknowledges support from the Wolfson Research Merit Award program of the U.K. Royal Society. YC acknowledges
the support of the Royal Society through a University Research Fellowship and an Enhancement Award.
This work used the DiRAC@Durham facility managed by the Institute for Computational Cosmology on behalf of the STFC DiRAC HPC Facility. The equipment was funded by BEIS capital funding via STFC capital grants ST/P002293/1, ST/R002371/1 and ST/S002502/1, Durham University and STFC operations grant ST/R000832/1. DiRAC is part of the National e-Infrastructure.

\section*{Data availability}
The raw \simba and IllustrisTNG simulation data and halo catalogues used in this paper are available at {\tt https://simba.roe.ac.uk} and {\tt https://www.tng-project.org/data}, respectively.
The tSZ data will be made available on request to the lead author.

\bibliographystyle{mnras}
\bibliography{draft}
\appendix

\section{Numerical convergence}\label{sec::convergence_test_full_physics_models}

As shown in Figure \ref{hgf_T_comp}, the difference of the in-halo $y$ contribution across models mainly arises from lowering the halo baryon fractions. For the \simba-50 `allphys' model and the \simba-100 model, although they have different box sizes and initial conditions, the same feedback mechanisms are implemented in both runs and the numerical resolutions are the same. Therefore, we can test the sensitivity of the predictions for $Y$--$M_{500}$ to the finite simulation volume by seeing whether these two `full physics' models converge well. 

\begin{figure}
\centering
    \includegraphics[width=.95\linewidth]{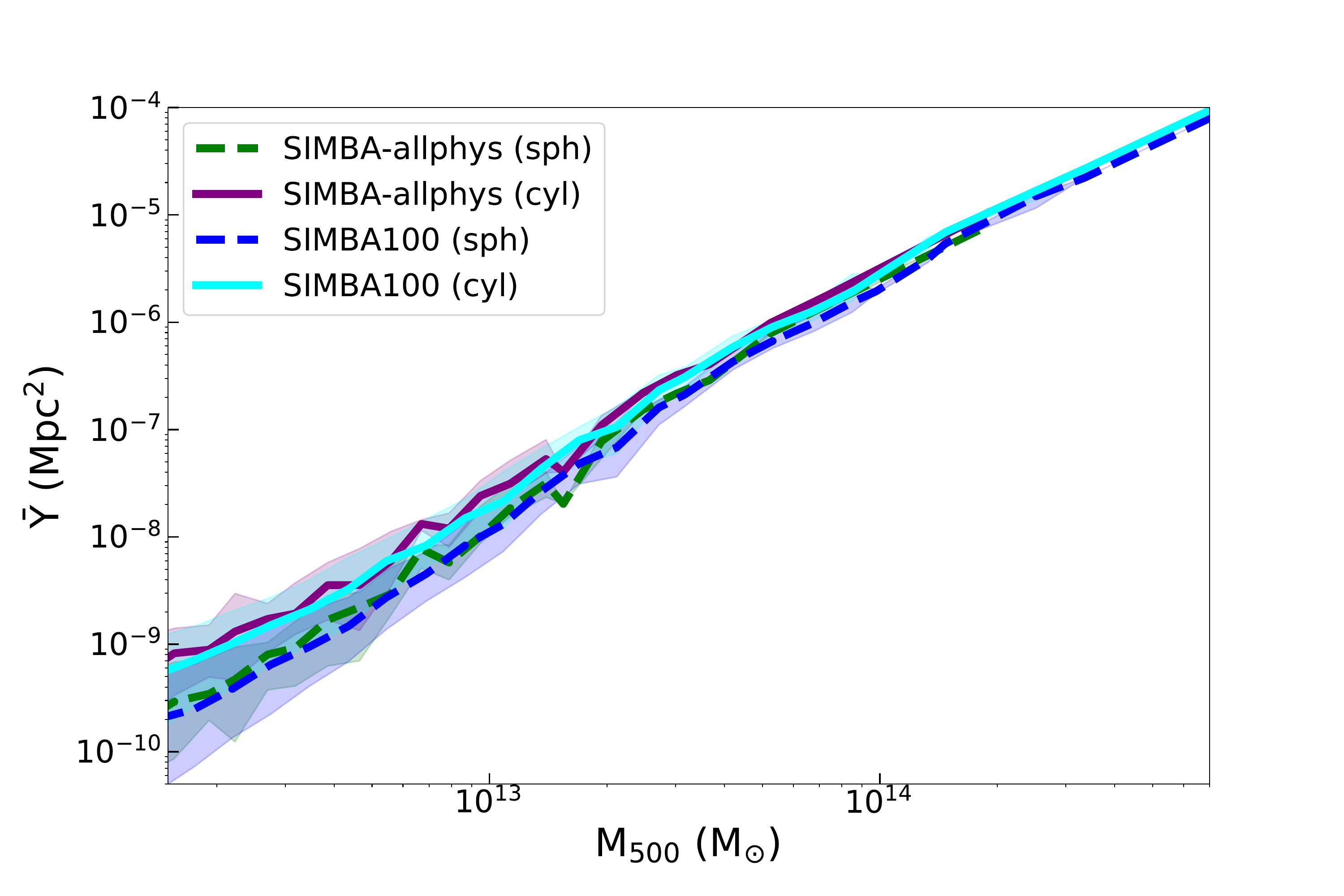}
\caption{Comparison of the measured $Y$--$M_{500}$ relation from the $\simba$50-`allphys' run and the $\simba$100 run at $z=0.17$.}
\label{Y_M_comp_simba_fullphys_models}
\end{figure}

Figure \ref{Y_M_comp_simba_fullphys_models} shows the resulting $Y$--$M_{500}$ relations measured for these two models at $z=0.17$. The shaded area shows the 68 percentile range around the median. It is clear that the curves measured from these two models are in very good agreement with each other, for both $Y_{\rm sph}$ (dashed lines) and $Y_{\rm cyl}$ (solid lines).  Hence our results are not sensitive to the simulation volume. Since \simba-100 has more haloes at high-mass end and is the main \simba volume, we only include the \simba-100 results in our main text, and do not show \simba-50 `allphys'.

\begin{figure}
\centering
    \includegraphics[width=.95\linewidth]{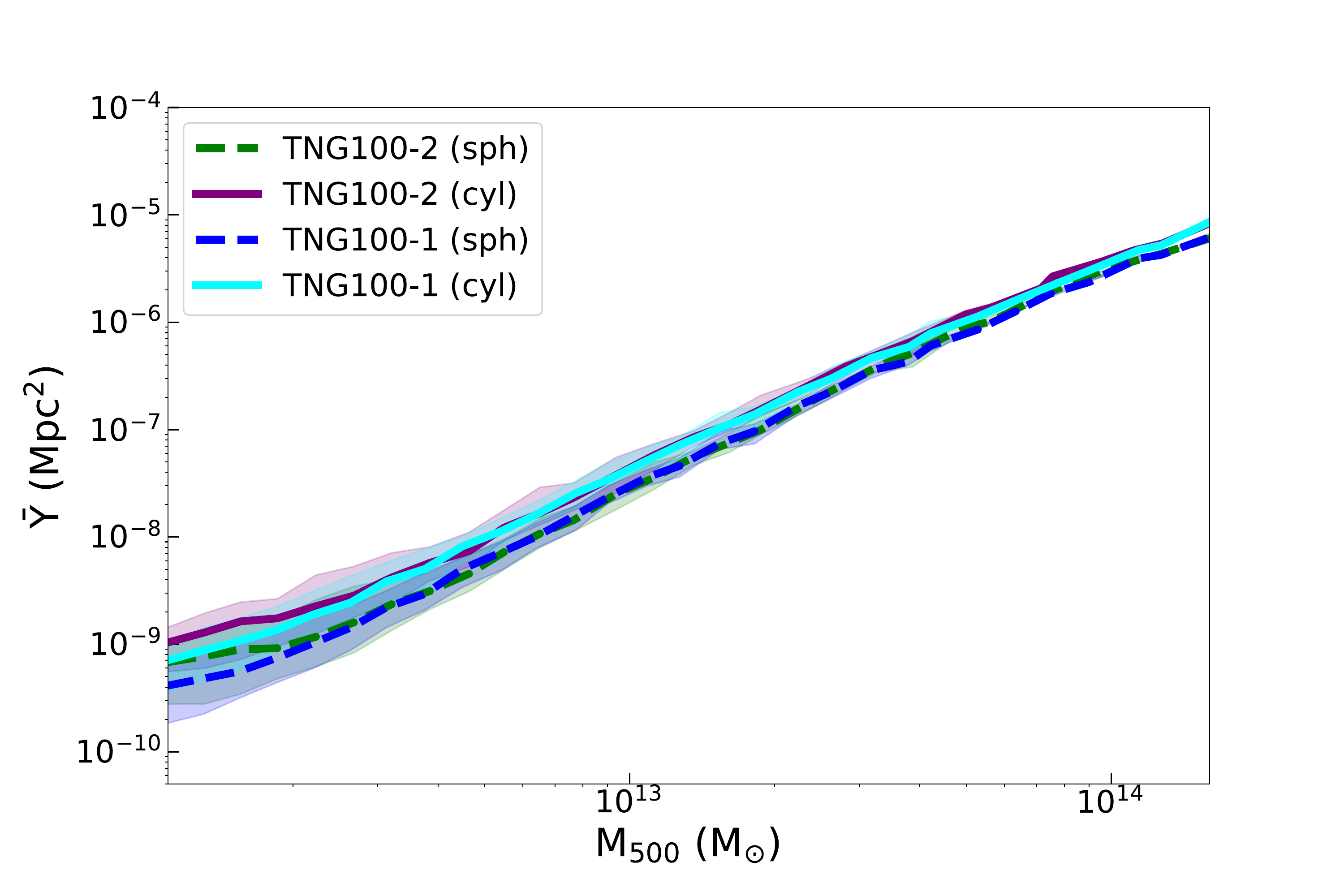}
\caption{Comparison of the measured $Y$--$M_{500}$ relation from the TNG100-2 and TNG100-1 runs at $z=0.17$.}
\label{Y_M_comp_tng_models}
\end{figure}

We further test resolution convergence using the IllustrisTNG runs.  Figure~\ref{Y_M_comp_tng_models} shows the $Y$--$M_{500}$ relation for the highest resolution IllustrisTNG-100 simulation in their suite, TNG100-1, versus the one used in this paper, TNG100-2.  These resolutions differ by a factor of 8 in particle mass, and a factor of 2 in spatial resolution.  Again, little difference is seen between the two cases, for either $Y_{\rm sph}$ (dashed lines) or $Y_{\rm cyl}$ (solid lines). This shows that numerical resolution is not likely to play a large role in our results.  This makes sense: since the tSZ signal is coming from hot gas that is on scales much greater than the resolutions of the simulations that we have employed in this work.  Hence, even though we do not have the necessary runs to conduct a similar resolution test on 
\simba, we expect that those results will likewise be insensitive to numerical resolution.

\section{Redshift evolution of the $Y$--$M$ relation}\label{sec::redshift_evolution_of_Y_M}

\begin{figure*}
\centering
    \includegraphics[width=1.0\linewidth]{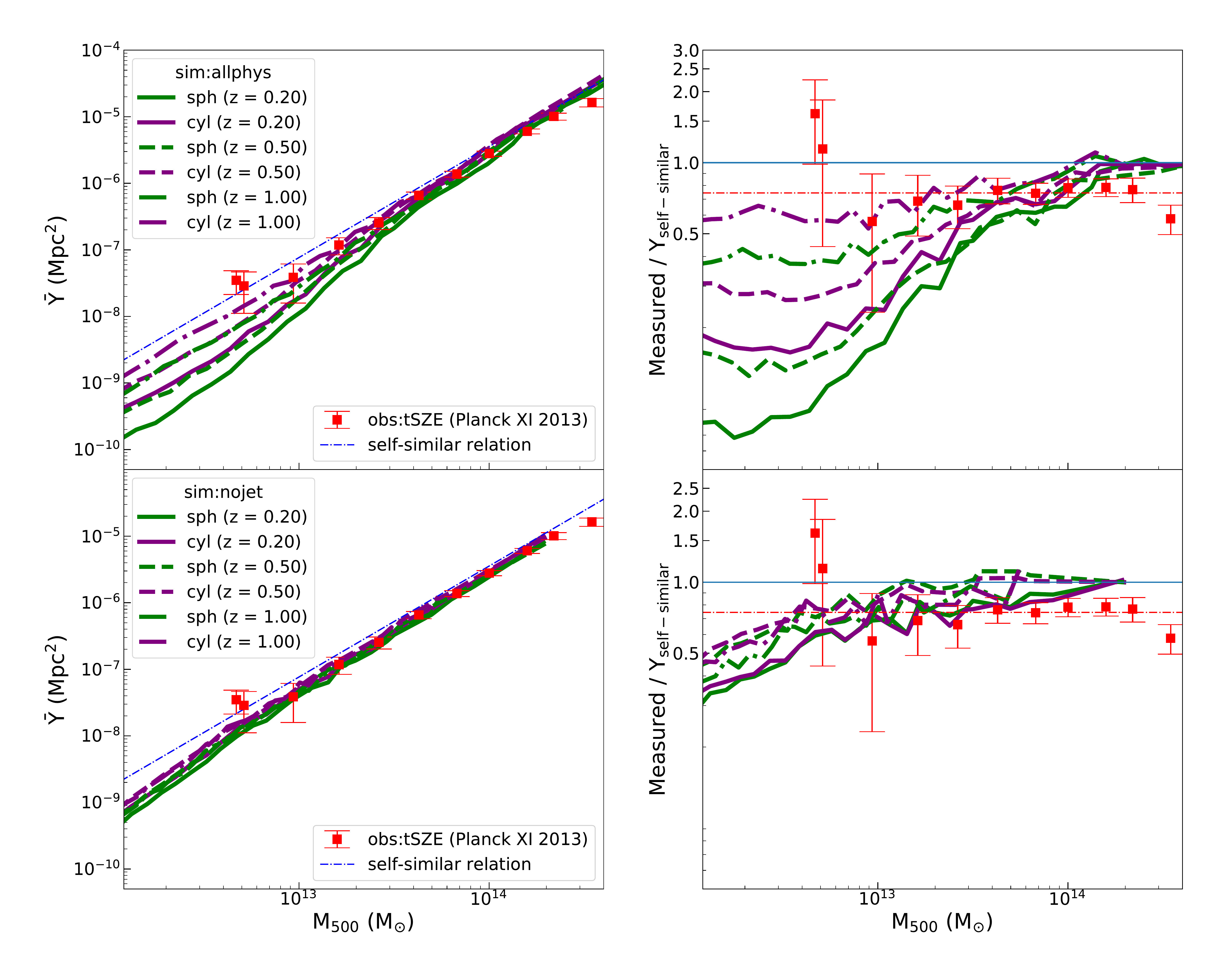}
\caption{\textit{Left}: comparison of the measured $Y$--$M_{500}$ relation for the \simba `allphys' and `nojet' model at different redshift values ($Y_{\rm sph}$, indicated by different linestyles in the legend. Green curves: averaging within a sphere; $Y_{\rm cyl}$; purple curves: projection along a cylinder, same as in Figure \ref{Y_M_relation_halo_mass}). \textit{Right}: ratio between the measured $Y$--$M$ relation and the self-similar model.}
\label{Y_M_comp_diff_zred}
\end{figure*}

While we have considered only a single redshift in the main analysis, for future experiments, it is interesting to examine how the tSZ signal evolves out to intermediate redshifts. Figure \ref{Y_M_comp_diff_zred} shows the resulting $Y_{\rm sph}$ and $Y_{\rm cyl}$ curves at $z=0.2$, $z=0.5$ and $z = 1.0$. $Y_{\rm sph}$ and $Y_{\rm cyl}$ have the same definition as discussed in Section \ref{sec::Y_M_relation}. 

From \simba `allphys', we see a tendency for a decreasing effect of AGN feedback with increasing redshift (upper panels in Figure~\ref{Y_M_comp_diff_zred}). This is expected as the main impact on large scales comes from \simba's jet feedback, which is tied primarily to the massive galaxy population becoming prominent at $z\la 1$. This drives the $Y$--$M$ relation to be asymptotically close to the self-similar model at high redshifts. 

The `nojet' model (lower panels in Figure~\ref{Y_M_comp_diff_zred}), in contrast, predicts virtually no redshift dependence for the $Y$--$M$ relation.  This demonstrates that the jet feedback is the primary driver of evolution.  Furthermore, it points towards the possibility of using the evolution of the $Y$--$M$ relation to test baryonic models.

Recently,  \citet{new_Y_M_paper} measured a slightly steeper $Y$--$M$ slope than the self-similar model for the tSZ signal by stacking the Planck tSZ $y$-signal around DESI galaxy clusters/groups. The slope is also reported to increase slightly with redshift. At face value, the redshift evolution seem contradictory to that predicted in \simba\ `allphys'. However, one needs to be careful given the significant differences in sample selection and in analysis methodology between the DESI work and that of the {\it Planck} analysis.  For instance, the DESI sample at higher redshifts is weighted towards more massive objects, and hence this measurements conflates the (opposite) effects of redshift evolution and halo mass dependence.  Given that we have found a strong dependence of the measured $Y$--$M$ relation on observational selection and analysis methodology, we leave a more careful comparison with \citet{new_Y_M_paper} for future work.

\label{lastpage}
\end{document}